\documentclass[showpacs]{revtex4} %twocolumn
\usepackage{epsfig,amsmath,amssymb,graphicx,bm,upgreek,textcomp}
\usepackage{natbib}
\usepackage[english]{babel} %

\begin{document}

\vspace{0mm}
\title{PHASE TRANSITION IN A PHONON GAS WITH PAIR CORRELATIONS} %
\author{Yu.M. Poluektov}
\email{yuripoluektov@kipt.kharkov.ua (y.poluekt52@gmail.com)} %
\affiliation{National Science Center ``Kharkov Institute of Physics and Technology'', 61108 Kharkov, Ukraine} %

\begin{abstract}
The phase transition to the state of a phonon gas with pairwise
correlations of interacting phonons with opposite momenta is
studied. A method for describing such phonon systems within the
framework of the self-consistent field model is developed and their
thermodynamic characteristics are calculated.   It is shown that a
phonon gas with pair correlations can exist in a state of unstable
thermodynamic equilibrium. The possibility of experimental
observation of a solid in such a phase is discussed.
\newline%
\end{abstract}
\maketitle

\section{Introduction}\vspace{-0mm} %\cite{}
A quantum mechanical description of a system of interacting bosons
by the second quantization method was first realized by Bogolyubov
\cite{R01}. He established the important role of correlations of
particles with opposite momenta. Accounting for such pair
correlations in Fermi systems later became the basis for
constructing the theory of superconductivity \cite{R02,R03}. The
role of pair correlations in Bose systems was studied in works
\cite{R04,R05,R06,R07,R08,R09}. Of undoubted interest is the
theoretical study of the role of similar correlations in a system of
massless particles (phonons). This problem is also relevant in
connection with experimental studies of the properties of quantum
crystals \cite{R10}.

The purpose of this work is a theoretical study of the possibility
of a phase transition of a gas of phonons in a solid into a state
with correlations of pairs of phonons with opposite momenta. A
general method for describing such states within the framework of
the self-consistent field model is developed. It is shown that in
the case of the existence of attraction between phonons, at a
certain critical temperature there becomes possible a second-order
phase transition to a state with correlations of phonon pairs, in
which the symmetry with respect to the phase transformation is
broken. The thermodynamic functions characterizing such a state are
calculated: entropy, pressure, energy, heat capacities,
thermodynamic coefficients. It is shown that, in contrast to the
standard theory of second-order phase transitions, the transition of
a phonon gas to the asymmetric state is accompanied not by a
decrease, but by an increase in the free energy. As a consequence,
the state with pair correlations of phonons is a state of unstable
thermodynamic equilibrium. The possibility of experimental detection
of such a state is discussed.

\section{Hamiltonian of interacting phonons}\vspace{-0mm} %
We choose the Hamiltonian that takes into account the interaction of
phonons with opposite momenta in the form analogous to the
Bardeen-Cooper-Schrieffer (BCS) Hamiltonian in the theory of
superconductivity \cite{R02}
\begin{equation} \label{01}
\begin{array}{l}
\displaystyle{%
   H=\frac{1}{2}\sum_k \hbar\omega_{\bf k} \Big(a_{\bf k}^+a_{\bf k} + a_{\bf -k}a_{\bf -k}^+ \Big) +   %
     \sum_{k,k'} U_{{\bf k}{\bf k}'} a_{\bf k}^+a_{\bf -k}^+a_{\bf -k'}a_{\bf k'},     %
}
\end{array}
\end{equation}
where $\omega_{\bf k}=\omega_{-\bf k}\equiv\omega_k=ck$, and the
operators $a_{\bf k}^+, a_{\bf k}$ are subject to the usual
commutation conditions for Bose particles. For simplicity, we will
use the scalar model, disregarding the polarization of phonons, and
assume that the phonon velocity $c$ is independent of temperature.
Accounting for polarization presents no fundamental difficulties.
The difference from the BCS theory is that the Hamiltonian (1)
describes a system of Bose particles, the number of which is not
fixed and is determined by temperature. We will study the phonon gas
in the model of a self-consistent field. The technique for studying
systems of interacting phonons in this approach with the fulfillment
of all thermodynamic relations was developed in works
\cite{R11,R12}. To formulate the model of a self-consistent field,
we break the total Hamiltonian (\ref{01}) into the sum of two terms
\begin{equation} \label{02}
\begin{array}{l}
\displaystyle{%
   H=H_S+H_C,  %
}
\end{array}
\end{equation}
where
\begin{equation} \label{03}
\begin{array}{l}
\displaystyle{%
   H_S=\frac{1}{2}\sum_k \Big[ \hbar\omega_{\bf k} \Big(a_{\bf k}^+a_{\bf k} + a_{\bf -k}a_{\bf -k}^+ \Big) +   %
       \Delta_{\bf k}a_{\bf k}^+a_{-\bf k}^+ + \Delta_{-\bf k}^*a_{-\bf k}a_{\bf k}\Big] + E_0,     %
}
\end{array}
\end{equation}
\vspace{-5mm} %
\begin{equation} \label{04}
\begin{array}{l}
\displaystyle{%
   H_C=\frac{1}{2}\sum_{k,k'} U_{{\bf k}{\bf k}'} a_{\bf k}^+a_{-\bf k}^+a_{-\bf k'}a_{\bf k'}-  %
       \frac{1}{2}\sum_k \Big[ \Delta_{\bf k}a_{\bf k}^+a_{-\bf k}^+ + \Delta_{-\bf k}^*a_{-\bf k}a_{\bf k}\Big]-E_0.    %
}%
\end{array}
\end{equation}
Let us note that, along with the phase-invariant operators $a_{\bf
k}^+a_{\bf k},\,a_{\bf -k}a_{\bf -k}^+$, the Hamiltonian (\ref{03})
also contains operators $a_{\bf k}^+a_{-\bf k}^+,\,a_{-\bf k}a_{\bf
k}$, leading to a violation of the symmetry of the state with
respect to the phase transformation $a_{\bf k}\rightarrow
e^{i\theta}a_{\bf k}$, $a_{\bf k}^+\rightarrow e^{-i\theta}a_{\bf
k}^+$, where $\theta$ is a real number. Within the framework of the
self-consistent field approximation, we will describe the phonon
system with the help of the approximating Hamiltonian $H_S$, which
includes so far unknown parameters $\Delta_{\bf k}=\Delta_{-\bf k}$
and $E_0$ which will be determined later. The effects associated
with the existence of the correlation Hamiltonian (\ref{04}) can be
taken into account using the perturbation theory \cite{R13,R14}, but
we restrict ourselves to the main approximation. The operators will
be averaged using the statistical operator
\begin{equation} \label{05}
\begin{array}{l}
\displaystyle{%
   \rho=\exp\beta\big(F-H_S\big),  %
}
\end{array}
\end{equation}
where $\beta=1/T$ is the inverse temperature and $F$ is the
normalization constant, the meaning of which will be clarified
below. The average of an arbitrary operator is $\big\langle
A\big\rangle= {\rm Sp}\rho A$, where the trace is taken over all
possible states in the self-consistent field approximation.

The Hamiltonian (\ref{03}) is quadratic in the phonon creation and
annihilation operators and, as a consequence, can be reduced to a
diagonal form using the canonical Bogolyubov transformation \cite{R01} %
\begin{equation} \label{06}
\begin{array}{l}
\displaystyle{%
   a_{\bf k}=u_{\bf k}\gamma_{\bf k}+\upsilon^*_{\bf k}\gamma^+_{-\bf k}, \qquad   %
   a_{\bf k}^+=u_{\bf k}^*\gamma_{\bf k}^+ +\upsilon_{\bf k}\gamma_{-\bf k},   %
}
\end{array}
\end{equation}
while the condition
\begin{equation} \label{07}
\begin{array}{l}
\displaystyle{%
   |u_{\bf k}|^2-|\upsilon_{\bf k}|^2=1   %
}
\end{array}
\end{equation}
ensures the fulfillment of the correct commutation relationships for
new operators $\gamma_{\bf k}^+, \gamma_{\bf k}$. In order for
off-diagonal terms to drop out in the Hamiltonian $H_S$, it is
necessary the fulfillment of the condition
\begin{equation} \label{08}
\begin{array}{l}
\displaystyle{%
   2\hbar\omega_ku_{\bf k}\upsilon_{\bf k}+\Delta_{\bf k}^*u_{\bf k}^2+\Delta_{\bf k}\upsilon_{\bf k}^2=0.   %
}
\end{array}
\end{equation}
This leads to the system of linear homogeneous equations
\begin{equation} \label{09}
\begin{array}{ccc}
\displaystyle{%
   \big(\hbar\omega_k-\varepsilon_{\bf k}\big)\upsilon_{\bf k}+\Delta_{\bf k}^*u_{\bf k}=0, %
}\vspace{3mm}\\ %
\displaystyle{%
  \Delta_{\bf k}\upsilon_{\bf k}+ \big(\hbar\omega_k+\varepsilon_{\bf k}\big)u_{\bf k}=0, %
}%
\end{array}
\end{equation}
and therefore
\begin{equation} \label{10}
\begin{array}{ccc}
\displaystyle{%
   \varepsilon_{\bf k}=\sqrt{\big(\hbar\omega_k\big)^2-\big|\Delta_{\bf k}\big|^2}. %
}%
\end{array}
\end{equation}
Taking into account condition (\ref{07}), we find
\begin{equation} \label{11}
\begin{array}{ccc}
\displaystyle{%
   \big|u_{\bf k}\big|^2=\frac{1}{2}\left(\frac{\hbar\omega_k}{\varepsilon_{\bf k}}+1\right), \qquad %
   \big|\upsilon_{\bf k}\big|^2=\frac{1}{2}\left(\frac{\hbar\omega_k}{\varepsilon_{\bf k}}-1\right),  %
}\vspace{3mm}\\ %
\displaystyle{%
  u_{\bf k}\upsilon_{\bf k}^*=-\frac{\Delta_{\bf k}}{2\varepsilon_{\bf k}}. %
}%
\end{array}
\end{equation}
As a result, the Hamiltonian (\ref{03}) takes the form of the
Hamiltonian of free phonons with energy $\varepsilon_{\bf k}$ (10): %
\begin{equation} \label{12}
\begin{array}{ccc}
\displaystyle{%
   H_S=\sum_k \varepsilon_{\bf k}\gamma_{\bf k}^+\gamma_{\bf k} + \frac{1}{2}\sum_k\varepsilon_{\bf k} + E_0. %
}%
\end{array}
\end{equation}
Here, the second term describes zero oscillations, and the quantity
$E_0$ characterizes the shift in the energy of the ground state upon
transition to a description in terms of new operators. The phonons
described by the operators $a_{\bf k}, a_{\bf k}^+$ will be called
``bare'' phonons. The phonons described by the operators
$\gamma_{\bf k}, \gamma_{\bf k}^+$   will be called ``dressed''
phonons. The distribution function of the ``dressed'' phonons has
the form of the Planck function
\begin{equation} \label{13}
\begin{array}{ccc}
\displaystyle{%
   f_{\bf k}\equiv \big\langle\gamma_{\bf k}^+\gamma_{\bf k}\big\rangle=\frac{1}{e^{\beta\varepsilon_{\bf k}}-1},  %
}%
\end{array}
\end{equation}
and their anomalous averages are equal to zero:
$\big\langle\gamma_{\bf k}^+\gamma_{-\bf
k}^+\big\rangle=\big\langle\gamma_{-\bf k}\gamma_{\bf
k}\big\rangle=0$. The total number of the ``dressed'' phonons is
\begin{equation} \label{14}
\begin{array}{ccc}
\displaystyle{%
   N_{ph}=\sum_k\frac{1}{e^{\beta\varepsilon_{\bf k}}-1}.  %
}%
\end{array}
\end{equation}
The pair averages for the ``bare'' phonons are given by the
relations:
\begin{equation} \label{15}
\begin{array}{ccc}
\displaystyle{%
   \big\langle a_{\bf k}^+ a_{-\bf k}^+\big\rangle=-\frac{\Delta_{\bf k}^*}{2\varepsilon_{\bf k}}\big(1+2f_{\bf k}\big), \qquad  %
   \big\langle a_{-\bf k} a_{\bf k}\big\rangle=-\frac{\Delta_{\bf k}}{2\varepsilon_{\bf k}}\big(1+2f_{\bf k}\big),   %
}\vspace{3mm}\\ %
\displaystyle{%
  \big\langle a_{\bf k}^+ a_{\bf k}\big\rangle=\frac{\hbar\omega_k}{2\varepsilon_{\bf k}}\big(1+2f_{\bf k}\big)-\frac{1}{2}.  %
}%
\end{array}
\end{equation}
The total number of the ``bare'' phonons is determined by the
formula
\begin{equation} \label{16}
\begin{array}{ccc}
\displaystyle{%
    N_{ph}^{(0)}=\sum_{{\bf k}}\big\langle a_{\bf k}^+ a_{\bf k}\big\rangle =%
    \frac{1}{2}\sum_{{\bf k}}\bigg[\frac{\hbar\omega_k}{\varepsilon_{\bf k}}\big(1+2f_{\bf k}\big)-1\bigg]. %
}%
\end{array}
\end{equation}
At $T\rightarrow 0$, the total number of the ``dressed'' phonons
$N_{ph}$ tends to zero, while the total number of the ``bare''
phonons $N_{ph}^{(0)}$ remains finite. The temperature dependences
of the phonon numbers are shown in Fig.\,4.

As we can see, along with the phase-invariant ``normal'' average
$\big\langle a_{\bf k}^+ a_{\bf k}\big\rangle$, there are also
phase-noninvariant ``anomalous'' averages $\big\langle a_{\bf k}^+
a_{-\bf k}^+\big\rangle$, $\big\langle a_{-\bf k} a_{\bf
k}\big\rangle$, which under phase transformations of operators are
transformed as follows: $\big\langle a_{-\bf k} a_{\bf k}\big\rangle
\rightarrow e^{2i\theta}\big\langle a_{-\bf k} a_{\bf
k}\big\rangle$,\, $\big\langle a_{\bf k}^+ a_{-\bf k}^+\big\rangle
\rightarrow e^{-2i\theta}\big\langle a_{\bf k}^+ a_{-\bf
k}^+\big\rangle$. With this in mind, it is obvious that the operator
of the observed energy quantity (\ref{03}) does not depend on the
choice of phase and on the whole remains invariant with respect to
phase transformations. The next task is to study the state of the
phonon gas, in which the ``anomalous'' averages that violate the
phase symmetry are nonzero.

\section{Parameters of the self-consistent Hamiltonian  }\vspace{-0mm} %
Let us move on to finding the parameters $\Delta_{\bf k}$ and $E_0$,
which enter into the approximating Hamiltonian (\ref{03}). The
normalization condition of the statistical operator (\ref{05}) ${\rm
Sp} \rho=1$ determines the normalization constant
\begin{equation} \label{17}
\begin{array}{ccc}
\displaystyle{%
    F=-T\ln{\rm Sp} e^{-\beta H_S}.
}%
\end{array}
\end{equation}
Taking into account the definition of entropy $S=-{\rm
Sp}\rho\ln\rho$, we find
\begin{equation} \label{18}
\begin{array}{ccc}
\displaystyle{%
    F=\big\langle H_S\rangle-TS,
}%
\end{array}
\end{equation}
whence it follows that the quantity $F$ has the meaning of free
energy in the self-consistent field model. The parameter $E_0$ is
determined from the requirement that the approximating Hamiltonian
(\ref{03}) be as close as possible to the initial Hamiltonian
(\ref{01}), which can be formulated as $\big\langle
(H-H_S)\big\rangle=0$ or $\big\langle H_C\rangle=0$, which gives
\begin{equation} \label{19}
\begin{array}{ccc}
\displaystyle{%
    E_0=\sum_{k,k'}U_{{\bf k}{\bf k}'}\big\langle a_{\bf k}^+a_{-\bf k}^+a_{-\mathbf{k}'}a_{\mathbf{k}'}\big\rangle %
    -\frac{1}{2}\sum_k\Big[\Delta_{\bf k}\big\langle a_{\bf k}^+a_{-\bf k}^+\big\rangle %
                            + \Delta_{-\bf k}^*\big\langle a_{-\bf k}a_{\bf k}\big\rangle \Big]. %
}%
\end{array}
\end{equation}
The average of four operators is expressed in terms of the products
of the averages of pairs of operators
\begin{equation} \label{20}
\begin{array}{ccc}
\displaystyle{%
    \big\langle a_{\bf k}^+a_{-\bf k}^+a_{-\mathbf{k}'}a_{\mathbf{k}'}\big\rangle = %
    \big\langle a_{\bf k}^+a_{-\bf k}^+\big\rangle\big\langle a_{-\mathbf{k}'}a_{\mathbf{k}'}\big\rangle + %
    \big\langle a_{\bf k}^+a_{-\mathbf{k}'}\big\rangle\big\langle a_{\bf k}^+a_{-\mathbf{k}'}\big\rangle +%
    \big\langle a_{\bf k}^+a_{\mathbf{k}'}\big\rangle\big\langle a_{-\bf k}^+a_{-\mathbf{k}'}\big\rangle. %
}%
\end{array}
\end{equation}
Since we are interested in the effects associated with the existence
of anomalous averages in the chosen model, we will take into account
in expansion (\ref{20}) the contribution of only the first term.
Then, taking into account (\ref{15}), we obtain
\begin{equation} \label{21}
\begin{array}{ccc}
\displaystyle{%
    E_0=\frac{1}{4}\sum_{k,k'}U_{{\bf k}{\bf k}'}\frac{\Delta_{\bf k}^*\Delta_{\bf k'}}{\varepsilon_{\bf k}\varepsilon_{\bf k'}} %
    \big(1+2f_{\bf k}\big)\big(1+2f_{\bf k'}\big) %
    +\frac{1}{2}\sum_k\frac{\big|\Delta_{\bf k}\big|^2}{\varepsilon_{\bf k}}\big(1+2f_{\bf k}\big). %
}%
\end{array}
\end{equation}
Taking into account the form of the Hamiltonian (\ref{12}), we find
the formula for the free energy
\begin{equation} \label{22}
\begin{array}{ccc}
\displaystyle{%
    F=E_0+\frac{1}{2}\sum_k\varepsilon_{\bf k}-T\ln {\rm Sp}\,e^{-\beta\sum_k\varepsilon_{\bf k}\gamma_{\bf k}^+\gamma_{\bf k}} . %
}%
\end{array}
\end{equation}
Here, the last term describes the free energy of an ideal gas of
``dressed'' phonons:
\begin{equation} \label{23}
\begin{array}{ccc}
\displaystyle{%
    F_0=-T\ln {\rm Sp}\,e^{-\beta\sum_k\varepsilon_{\bf k}\gamma_{\bf k}^+\gamma_{\bf k}}= %
    T\sum_k\ln\big(1-e^{-\beta\varepsilon_{\bf k}}\big). %
}%
\end{array}
\end{equation}
Comparing formulas (\ref{12}), (\ref{13}), (\ref{18}) and taking
into account the definition of entropy, we find for entropy a
natural expression in terms of the distribution function of
``dressed'' phonons:
\begin{equation} \label{24}
\begin{array}{ccc}
\displaystyle{%
    S=\sum_k\Big[\big(1+f_{\bf k}\big)\ln \big(1+f_{\bf k}\big)-f_{\bf k}\ln f_{\bf k}\Big]. %
}%
\end{array}
\end{equation}
Since in the limit $T\rightarrow 0$, the distribution function
(\ref{13}) tends to zero, then the entropy, as required, also tends
to zero.

Since the dispersion law for ``bare'' phonons $\omega_k=ck$ is
linear in the wave number $k$, then it is natural to assume, and
this is justified by the result, that the same dependence is also
preserved for the parameter $\Delta_{\bf k}$, so we set
\begin{equation} \label{25}
\begin{array}{ccc}
\displaystyle{%
    \Delta_{\bf k}=\sigma\hbar ck=\sigma\hbar\omega_k, %
}%
\end{array}
\end{equation}
where $\sigma$ has the meaning of the order parameter in the
asymmetric state, which does not depend on the wave number of the
photon. Then it turns out that the energy of the ``dressed'' photon (10) %
\begin{equation} \label{26}
\begin{array}{ccc}
\displaystyle{%
    \varepsilon_{\bf k}=\xi\!\cdot\!\hbar\omega_k=\hbar\omega_k\sqrt{1-|\sigma|^2}, \qquad %
    \xi\equiv\sqrt{1-|\sigma|^2}
}%
\end{array}
\end{equation}
decreases with increasing $|\sigma|$. In this case, formulas
(\ref{11}) will take the form
\begin{equation} \label{27}
\begin{array}{ccc}
\displaystyle{%
   \big|u_{\bf k}\big|^2\equiv \big|u\big|^2 =\frac{1}{2}\left(\frac{1}{\xi}+1\right), \qquad %
   \big|\upsilon_{\bf k}\big|^2\equiv\big|\upsilon\big|^2  =\frac{1}{2}\left(\frac{1}{\xi}-1\right),  %
}\vspace{3mm}\\ %
\displaystyle{%
  u_{\bf k}\upsilon_{\bf k}^*\equiv u\upsilon^*=-\frac{\sigma}{2\xi}. %
}%
\end{array}
\end{equation}

We will describe the interaction between phonons with the help of
one constant $g_0$ , assuming
\begin{equation} \label{28}
\begin{array}{ccc}
\displaystyle{%
    U_{{\bf k}{\bf k}'}=\frac{g_0}{V},
}%
\end{array}
\end{equation}
where $V$ is the volume. We will not specify the nature of the
interaction between phonons, assuming that the effective interaction
between them can be caused by nonlinear effects. Then the energy of
the ground state (\ref{21})
\begin{equation} \label{29}
\begin{array}{ccc}
\displaystyle{%
    E_0=\frac{g_0}{4V}\frac{|\sigma|^2}{\xi^2} \bigg(\sum_k\big(1+2f_{\bf k}\big)\bigg)^{\!2}+ %
    \frac{\hbar c}{2}\frac{|\sigma|^2}{\xi}\sum_k k\big(1+2f_{\bf k}\big), %
}%
\end{array}
\end{equation}
and free energy (\ref{22}) can be written in the form
\begin{equation} \label{30}
\begin{array}{ccc}
\displaystyle{%
    F=\frac{g_0}{4V}\frac{|\sigma|^2}{\xi^2} I_2^2 + %
    \frac{\hbar c}{2}\frac{|\sigma|^2}{\xi}I_3 + \frac{\hbar c}{2}\,\xi\!\cdot\!I_1 - T\!\cdot\!I_4. %
}%
\end{array}
\end{equation}
Here $T$ is temperature, and
\begin{equation} \label{31}
\begin{array}{ccc}
\displaystyle{%
    I_0\equiv\sum_k 1, \quad %
    I_1\equiv\sum_k k, \quad %
    I_2\equiv\sum_k \big(1+2f_{\bf k}\big), \quad %
    I_3\equiv\sum_k k\big(1+2f_{\bf k}\big), \quad %
    I_4\equiv\sum_k \ln\!\big(1+f_{\bf k}\big). %
}%
\end{array}
\end{equation}
These quantities can be calculated using the transition in
(\ref{31}) from summation to integration. Integration is carried out
from zero to the Debye wave number $k_D$, which is determined by the
density of number of particles in a solid
\begin{equation} \label{32}
\begin{array}{ccc}
\displaystyle{%
    n\equiv\frac{N}{V}=\frac{k_D^3}{6\pi^2}.
}%
\end{array}
\end{equation}
The calculation results are expressed in terms of the Debye
functions
\begin{equation} \label{33}
\begin{array}{ccc}
\displaystyle{%
    D_n(x)=\frac{n}{x^n}\int_0^x\frac{z^ndz}{e^z-1}, \qquad (n\geq 1),    %
}%
\end{array}
\end{equation}
and related to them functions
\begin{equation} \label{34}
\begin{array}{ccc}
\displaystyle{%
    \Phi_n(x)=1+\frac{2(1+n)}{n}\frac{D_n(x)}{x}. %
}%
\end{array}
\end{equation}
Some properties of these functions are given in Appendix A. The
calculation gives:
\begin{equation} \label{35}
\begin{array}{ccc}
\displaystyle{%
   I_0=\frac{Vk_D^3}{6\pi^2}\equiv N, \quad %
   I_1=\frac{3}{4}Nk_D, \quad %
   I_2=N\Phi_2\bigg(\frac{\xi}{\tau}\bigg), \quad %
   I_3=\frac{3}{4}Nk_D\Phi_3\bigg(\frac{\xi}{\tau}\bigg),  %
}\vspace{3mm}\\ %
\displaystyle{%
   I_4=\frac{N}{3}\bigg[D_3\bigg(\frac{\xi}{\tau}\bigg)-3\ln\!\Big(1-e^{-\frac{\xi}{\tau}}\Big)\bigg].  %
}%
\end{array}
\end{equation}
Here, the reduced dimensionless temperature is defined
\begin{equation} \label{36}
\begin{array}{ccc}
\displaystyle{%
    \tau\equiv\frac{T}{\Theta_D}, %
}%
\end{array}
\end{equation}
where $\Theta_D\equiv\hbar ck_D$ is Debye energy or Debye
temperature in energy units. In what follows, we will also use the
notation $\overline{\tau}\equiv 1/\tau$ for the reciprocal reduced
temperature. Taking into account relations (\ref{34}),\,(\ref{35}),
the free energy for the state with pair correlations takes the form
\begin{equation} \label{37}
\begin{array}{ccc}
\displaystyle{%
  \frac{F}{\Theta_DN}\equiv f(\xi,\tau)= \frac{\xi}{2}+\frac{3}{16}g\big(\xi^{-2}-1\big)\Phi_2^2\big(\xi\overline{\tau}\big)+ %
  \frac{3}{8}\Big(\xi^{-1}-\frac{4}{3}\xi\Big)\Phi_3\big(\xi\overline{\tau}\big)+ %
  \tau\ln\!\big(1-e^{-\xi\overline{\tau}}\big), %
}%
\end{array}
\end{equation}
where the notation for the dimensionless interaction constant of
phonons is used
\begin{equation} \label{38}
\begin{array}{ccc}
\displaystyle{%
  g=\frac{4}{3}\frac{g_0n}{\Theta_D}. %
}%
\end{array}
\end{equation}
Thermodynamic potentials, which, along with equilibrium values,
additionally include arbitrary parameters, in our case it is
$\sigma$ or $\xi\equiv\sqrt{1-|\sigma|^2}$, are potentials with
incomplete thermodynamic equilibrium. The equilibrium value of the
parameter is found from the condition of the extremum of such a
thermodynamic potential
\begin{equation} \label{39}
\begin{array}{ccc}
\displaystyle{%
  \frac{\partial F}{\partial|\sigma|}=-\frac{\partial F}{\partial \xi}\frac{|\sigma|}{\xi}=0. %
}%
\end{array}
\end{equation}
This equation has two solutions. The first solution $\sigma=0$
corresponds to the normal phase, for which the Debye theory is valid
and the free energy is given by
\begin{equation} \label{40}
\begin{array}{ccc}
\displaystyle{%
  \frac{F_D}{\Theta_DN}\equiv f_D(\tau)= \frac{1}{2}- \frac{1}{8}\Phi_3\big(\overline{\tau}\big)+ %
  \tau\ln\!\big(1-e^{-\overline{\tau}}\big). %
}%
\end{array}
\end{equation}
We will also refer to this normal state as the Debye phase. The
second solution, where $\partial F/\partial\xi=0$ and $\sigma\neq
0$, corresponds to the phase with broken phase symmetry, in which
the correlations of phonons with opposite momenta are nonzero. The
condition $\partial F/\partial\xi=\partial f/\partial\xi=0$ leads to
the equation
%\begin{equation} \label{41}
%\begin{array}{ccc}
%\displaystyle{%
%  \frac{3}{4}g\bigg[\big(\xi^{-2}-1\big)\,{\rm cth}\frac{\xi}{2\tau}-\! %
%  \Big(\,\frac{4}{3}\,\xi^{-2}-1\Big)\Phi_2\big(\xi\overline{\tau}\big)\bigg]\Phi_2\big(\xi\overline{\tau}\big)\,+ %
%}\vspace{3mm}\\ %
%\displaystyle{%
%   +\,\xi\bigg[\big(\xi^{-2}-1\big)\,{\rm cth}\frac{\xi}{2\tau}-\! %
%   \Big(\,\frac{5}{4}\,\xi^{-2}-1\Big)\Phi_3\big(\xi\overline{\tau}\big)\bigg]=0. %
%}%
%\end{array}
%\end{equation}
\begin{equation} \label{41}
\begin{array}{ccc}
\displaystyle{%
  \frac{3}{4}\,g\bigg[\big(\xi^{-2}-1\big)\,{\rm cth}\frac{\xi}{2\tau}-\! %
  \Big(\,\frac{4}{3}\,\xi^{-2}-1\Big)\Phi_2\big(\xi\overline{\tau}\big)\bigg]\Phi_2\big(\xi\overline{\tau}\big)+ %
  \xi\bigg[\big(\xi^{-2}-1\big)\,{\rm cth}\frac{\xi}{2\tau}-\! %
   \Big(\,\frac{5}{4}\,\xi^{-2}-1\Big)\Phi_3\big(\xi\overline{\tau}\big)\bigg]=0. %
}%
\end{array}
\end{equation}
This equation together with formula (\ref{37}) in parametric form
determines the temperature dependence of the equilibrium free energy
in the asymmetric phase. The order parameter can vary within limits
$\xi_0\leq\xi\leq 1$. Equation (\ref{41}) has a solution only for a
negative interaction constant $g=-|g|$, such that $|g|<1$. At zero
temperature, as follows from (\ref{41}), we have $\xi_0=|g|$. The
transition from the asymmetric phase, where $\sigma\neq 0$, to the
symmetric phase occurs when the order parameter vanishes, $\sigma=0$
or $\xi=1$. Thus, the phase transition temperature
$\tau_c=T_c/\Theta_D$ is determined by the equation
\begin{equation} \label{42}
\begin{array}{ccc}
\displaystyle{%
  |g|\Phi_2^2(\overline{\tau}_c\big)-\Phi_3(\overline{\tau}_c\big)=0, %
}%
\end{array}
\end{equation}
where $\overline{\tau}_c=1/\tau_c$. The dependence of the critical
temperature $\tau_c$ on the modulus of the interaction constant
$|g|$ is shown in Figure 1. The unusual character of this dependence
is noteworthy. As we can see, the critical temperature increases
with decreasing the modulus of the interaction constant, and when
$|g|=1$ the critical temperature turns to zero. It would seem that,
by decreasing the interaction we can thereby increase the phase
transition temperature, but, as will be shown below, with decreasing
$|g|$ the stability region of the asymmetric phase narrows.  It
should be emphasized that the states of free phonons and systems of
phonons with pair correlations are qualitatively different due to
their different symmetries. Since the dependence of thermodynamic
quantities on the interaction constant in the asymmetric phase is
non-analytical, it is impossible to pass to the limit
$|g|\rightarrow 0$. Therefore, the asymmetric phase does not pass
into the symmetric phase when the interaction is weakened. This
situation is typical for all systems with broken phase symmetry \cite{R15}. %
\vspace{0mm} %
\begin{figure}[h!]
\vspace{-0mm}  \hspace{0mm}
\includegraphics[width = 7.45cm]{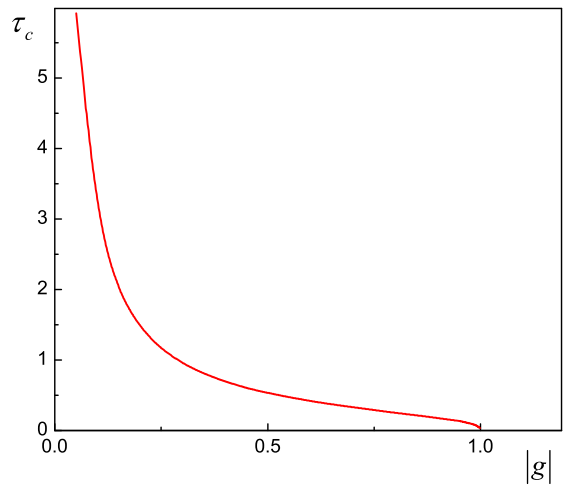} % 1.0\columnwidth
\vspace{-3mm} %
\caption{\label{fig01} %\hspace{10mm} %
Dependence of the  critical temperature $\tau_c=T_c/\Theta_D$ on the
interaction constant $|g|$.
}%
\end{figure}

Formulas (\ref{37}) and (\ref{41}) make it possible to construct the
temperature dependence of the equilibrium free energy, which is
shown in Fig. 2. This figure also shows the temperature dependence
(\ref{40}) of the free energy of the Debye phase.
\vspace{0mm} %
\begin{figure}[h!]
\vspace{-0mm}  \hspace{0mm}
\includegraphics[width = 8.25cm]{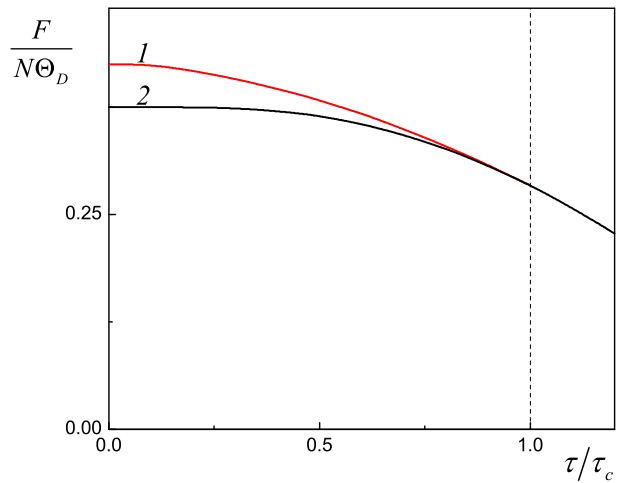} % 1.0\columnwidth
\vspace{-3mm} %
\caption{\label{fig02} %\hspace{10mm} %
The free energies in the asymmetric ({\it 1}) and symmetric ({\it
2}) states at $|g|=0.6$.
}%
\end{figure}

For the asymmetric phase to be thermodynamically stable, the
extremum of the free energy must be a minimum. In present case, this
condition is not satisfied at all temperatures, since
$\partial^2f\big/\partial\xi^2<0$ and the free energy of the
asymmetric phase is always greater than the free energy of the
symmetric phase. Thus, the state with pair correlations of phonons
is a state with unstable thermodynamic equilibrium. The possibility
of experimental observation of such a state will be discussed below.

\section{Thermodynamic quantities}\vspace{-0mm} %
Although the equilibrium state of phonons with pair correlations
corresponds not to the minimum, but to the maximum of the free
energy, and therefore is a state with an unstable equilibrium,
nevertheless, let us analyze the thermodynamic properties of such a
state. Having expressions (\ref{37}),\,(\ref{40}) for the
equilibrium free energy, we can calculate all thermodynamic
quantities in the asymmetric phase. We first write down the free
energy differential
\begin{equation} \label{43}
\begin{array}{ccc}
\displaystyle{%
  dF=\bigg(\frac{\partial F}{\partial\xi}\bigg)_{\!T,V}d\xi + %
     \bigg(\frac{\partial F}{\partial T}\bigg)_{\!V,\xi}dT + %
     \bigg(\frac{\partial F}{\partial V}\bigg)_{\!T,\xi}dV.  %
}%
\end{array}
\end{equation}
When condition (\ref{41}) is satisfied, the first term in (\ref{43})
is equal to zero. Taking into account the general thermodynamic
identities $dF=-SdT-pdV$ and $E=F+TS$, we find that the entropy,
pressure and energy in the self-consistent field model are
determined by the usual formulas
\begin{equation} \label{44}
\begin{array}{ccc}
\displaystyle{%
  S=-\bigg(\frac{\partial F}{\partial T}\bigg)_{\!V,\xi}, \qquad %
  p=-\bigg(\frac{\partial F}{\partial V}\bigg)_{\!T,\xi}, \qquad %
  E=F-T\bigg(\frac{\partial F}{\partial T}\bigg)_{\!V,\xi}.  %
}%
\end{array}
\end{equation}
It should be emphasized that, due to (\ref{39}), here it is not
necessary to differentiate with respect to the parameter $\xi$.
Since we assumed that the speed of ``bare'' phonons does not depend
on temperature, then in this approximation the Debye energy is a
function of only density or, at a constant number of particles in a
solid $N$, a function of only volume $\Theta_D\equiv\Theta_D(V)$. To
characterize this dependence, the Gr\"{u}neisen parameter is
introduced
\begin{equation} \label{45}
\begin{array}{ccc}
\displaystyle{%
  \Gamma\equiv - \frac{V}{\Theta_D}\frac{\partial\Theta_D}{\partial V}= %
  \frac{n}{\Theta_D}\frac{\partial\Theta_D}{\partial n} =   %
  \frac{1}{3}+\frac{n}{c}\frac{dc}{dn}. %
}%
\end{array}
\end{equation}
If we neglect the dependence of the phonon velocity on density, then
$\Gamma=1/3$. If the phonon velocity depends on density in a
power-law manner $c\sim n^k$, then the Gr\"{u}neisen parameter in
this case is also just a number $\Gamma=1/3+k$. In the following, we
assume the Gr\"{u}neisen parameter to be a positive number
independent of temperature and density. This assumption is satisfied
with good accuracy in a wide temperature range \cite{R16}. All
thermodynamic characteristics can be expressed in terms of the
function $f(\xi,\tau)$ (\ref{37}) and its derivatives. Let us
introduce the definitions of the following functions of variables $\xi,\tau$: %
\begin{equation} \label{46}
\begin{array}{ccc}
\displaystyle{%
  \psi(\xi,\tau)\equiv\frac{\partial f}{\partial\xi}=\frac{1}{2}\,{\rm cth}\bigg(\frac{\xi}{2\tau}\bigg) %
  -\frac{3g}{8}\xi^{-3}\Phi_2^2\big(\xi\overline{\tau}\big) %
  +\frac{3g}{8}\big(\xi^{-2}-1\big)\Phi_2\big(\xi\overline{\tau}\big)\frac{1}{\tau}\,\dot{\Phi}_2\big(\xi\overline{\tau}\big)- %
}\vspace{3mm}\\ %
\displaystyle{%
   -\frac{3}{8}\Big(\xi^{-2}+\frac{4}{3}\,\Big)\Phi_3\big(\xi\overline{\tau}\big)  %
   +\frac{3}{8}\Big(\xi^{-1}-\frac{4}{3}\,\xi\,\Big)\frac{1}{\tau}\,\dot{\Phi}_3\big(\xi\overline{\tau}\big), %
}%
\end{array}
\end{equation}
\begin{equation} \label{47}
\begin{array}{ccc}
\displaystyle{%
  s(\xi,\tau)\equiv-\frac{\partial f}{\partial\tau}=\frac{3g}{8}\frac{\xi}{\tau^2}\big(\xi^{-2}-1\big)\Phi_2\big(\xi\overline{\tau}\big)\dot{\Phi}_2\big(\xi\overline{\tau}\big)+ %
  +\frac{3}{8}\frac{\xi}{\tau^2}\Big(\xi^{-1}-\frac{4}{3}\,\xi\,\Big)\dot{\Phi}_3\big(\xi\overline{\tau}\big) -%
}\vspace{3mm}\\ %
\displaystyle{%
   -\ln\!\big(1-e^{-\xi\overline{\tau}}\big)+\frac{\xi}{\tau}\frac{1}{e^{\xi\overline{\tau}}-1}, %
}%
\end{array}
\end{equation}
\vspace{-2mm}
\begin{equation} \label{48}
\begin{array}{ccc}
\displaystyle{%
  \varphi(\xi,\tau)\equiv f-\tau\frac{\partial f}{\partial\tau}=\frac{3}{2}\big(\xi^{-1}-\xi\big)\,{\rm cth}\bigg(\frac{\xi}{2\tau}\bigg)+ %
}\vspace{3mm}\\ %
\displaystyle{%
  +\frac{3g}{16}\big(\xi^{-2}-1\big)\Phi_2\big(\xi\overline{\tau}\big) %
  \bigg[6\,{\rm cth}\bigg(\frac{\xi}{2\tau}\bigg)-5\,\Phi_2\big(\xi\overline{\tau}\big)\bigg]   %
  -\frac{9}{8}\Big(\xi^{-1}-\frac{4}{3}\,\xi\,\Big)\Phi_3\big(\xi\overline{\tau}\big). %
}%
\end{array}
\end{equation}
Taking into account the condition $\psi(\xi,\tau)=0$ equivalent to
(\ref{41}), the corresponding equilibrium functions $s(\tau)$ and
$\varphi(\tau)$ can be calculated, which always take positive
values. The equilibrium values of entropy, pressure and energy are
expressed through these functions:
\begin{equation} \label{49}
\begin{array}{ccc}
\displaystyle{%
  \frac{S}{N}=s(\tau), \qquad  %
  p=\Gamma n\Theta_D\varphi(\tau), \qquad %
  E=N\Theta_D\varphi(\tau).  %
}%
\end{array}
\end{equation}
From (\ref{49}), in particular, it follows that pressure and energy
are related as
\begin{equation} \label{50}
\begin{array}{ccc}
\displaystyle{%
  p=\Gamma\frac{E}{V}. %
}%
\end{array}
\end{equation}
At $\Gamma=1/3$, this formula coincides with the equation of state
for a photon gas \cite{R17,R18}. The temperature dependences of
entropy, pressure and energy are shown in Fig.\,3.
\vspace{0mm} %
\begin{figure}[t!]
\vspace{-0mm}  \hspace{0mm}
\includegraphics[width = 15.70cm]{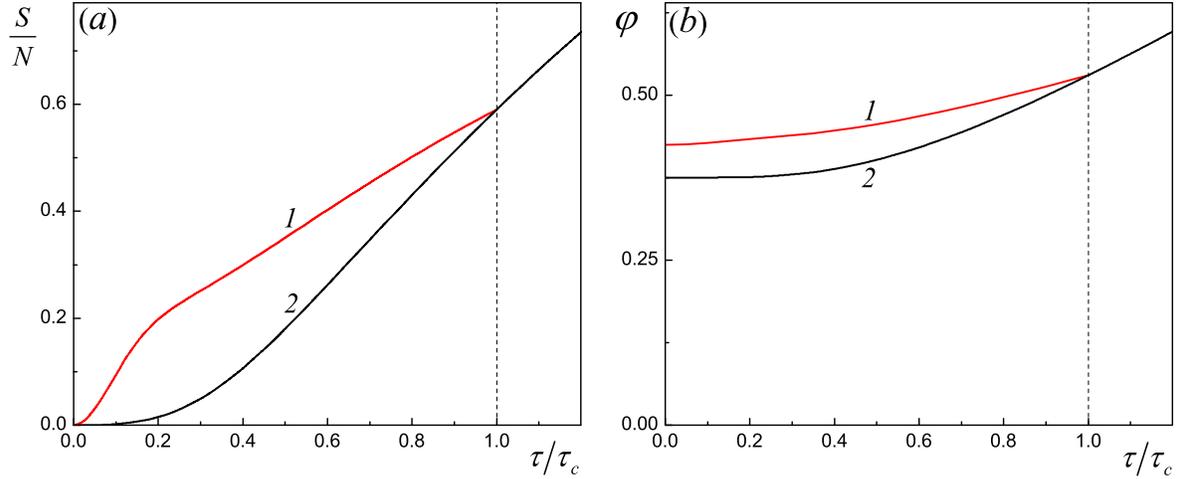} % 1.0\columnwidth
\vspace{-3mm} %
\caption{\label{fig03} %\hspace{10mm} %
Entropy $s=S/N$ ({\it a}), pressure and energy $\varphi=p\big/\Gamma
n\Theta_D=E\big/N\Theta_D$ ({\it b}) in the asymmetric ({\it 1}) and
symmetric ({\it 2}) states at $|g|=0.6$.
}%
\end{figure}

The numbers of ``dressed'' $N_{ph}$  and ``bare'' $N_{ph}^{(0)}$
phonons, according to (\ref{14}) and (\ref{16}), are expressed by
the formulas
\begin{equation} \label{51}
\begin{array}{ccc}
\displaystyle{%
  N_{ph}=\frac{3}{2}N\frac{\tau}{\xi}D_2\bigg(\frac{\xi}{\tau}\bigg), \qquad  %
  N_{ph}^{(0)}=\frac{N}{2}\bigg(\frac{1}{\xi}-1\bigg)+\frac{3}{2}N\frac{\tau}{\xi^2}D_2\bigg(\frac{\xi}{\tau}\bigg).   %
}%
\end{array}
\end{equation}
The temperature dependences of the number of phonons are shown in
Fig.\,4. As was noted, the number of ``dressed'' phonons, which
determine the temperature dependences of thermodynamic quantities,
also tends to zero as the temperature approaches zero (curve {\it 1}
in Fig.\,4). In this case, the number of ``bare'' phonons, which do
not contribute to entropy, remains finite and even increases (curve
{\it 2} in Fig.\,4).
\vspace{0mm} %
\begin{figure}[h!]
\vspace{-0mm}  \hspace{0mm}
\includegraphics[width = 7.9cm]{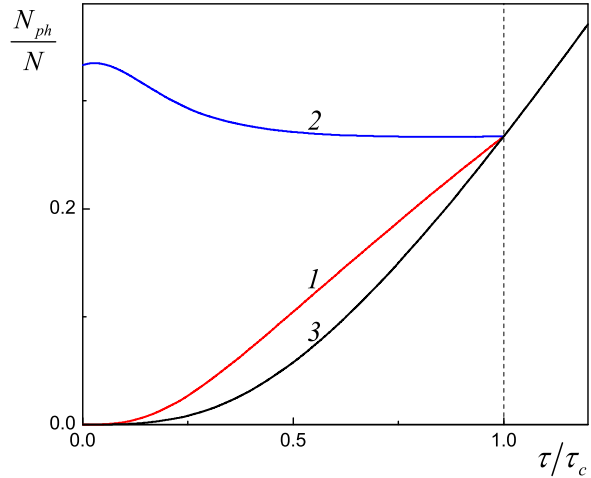} % 1.0\columnwidth
\vspace{-3mm} %
\caption{\label{fig04} %\hspace{10mm} %
Temperature dependencies of the number of phonons: $N_{ph}$ --
``dressed'' ({\it 1}); $N_{ph}^{(0)}$ -- ``bare'' ({\it 2});
$N_{ph}^{(D)}$  -- ``Debye'' ({\it 3}) phonons at $|g|=0.6$.
}%
\end{figure}

In order to calculate the heat capacities and thermodynamic
coefficients, one should find the entropy and pressure
differentials. Here it is necessary to take into account that, due
to condition $d\psi(\xi,\tau)=0$, the differentials of $\xi$ and
$\tau$ are related as follows
\begin{equation} \label{52}
\begin{array}{ccc}
\displaystyle{%
  d\xi=-\frac{\big(\partial\psi\big/\partial\tau\big)_\xi}{\big(\partial\psi\big/\partial\xi\big)_\tau}\,d\tau, %
}%
\end{array}
\end{equation}
where
\begin{equation} \label{53}
\begin{array}{ccc}
\displaystyle{%
  d\tau=\frac{dT}{\Theta_D}+\Gamma\frac{T}{\Theta_D}\frac{dV}{V}.   %
}%
\end{array}
\end{equation}
As a result, we get:
\begin{equation} \label{54}
\begin{array}{ccc}
\displaystyle{%
  \frac{dS}{N}=\tau\Omega\bigg(\frac{dT}{T}+\Gamma\frac{dV}{V}\bigg),  %
}%
\end{array}
\end{equation}
\vspace{-3mm}
\begin{equation} \label{55}
\begin{array}{ccc}
\displaystyle{%
  dp=\Gamma n\Omega T\frac{dT}{\Theta_D} + \Gamma\big(1+\Gamma\big)n\Theta_D\varphi\bigg[\frac{\Gamma}{\big(1+\Gamma\big)}\frac{\Omega\tau^2}{\varphi}-1\bigg]\frac{dV}{V}. %
}%
\end{array}
\end{equation}
Here the quantity
\begin{equation} \label{56}
\begin{array}{ccc}
\displaystyle{%
  \Omega=\Omega(\xi,\tau)\equiv\frac{\big(\partial^2f\big/\partial\xi\partial\tau\big)^2-\big(\partial^2f\big/\partial\xi^2\big)\big(\partial^2f\big/\partial\tau^2\big)}{\big(\partial^2f\big/\partial\xi^2\big)} %
}%
\end{array}
\end{equation}
is expressed in terms of the second derivatives of the function
$f(\xi,\tau)$ (\ref{37}), the explicit form of which is given in
Appendix B. Taking into account the condition $\psi(\xi,\tau)=0$,
the equilibrium function $\Omega(\tau)$ is calculated, through which
are expressed the heat capacities and thermodynamic coefficients.

{\it Heat capacities}. Assuming in formulas (\ref{54}),\,(\ref{55})
first $dV=0$, and then $dp=0$, we obtain expressions for the
isochoric and isobaric heat capacities:
\begin{equation} \label{57}
\begin{array}{ccc}
\displaystyle{%
  C_V=T\bigg(\frac{\partial S}{\partial T}\bigg)_{\!V}=\frac{NT\Omega}{\Theta_D},  %
}%
\end{array}
\end{equation}
\vspace{-4mm}
\begin{equation} \label{58}
\begin{array}{ccc}
\displaystyle{%
  C_p=T\bigg(\frac{\partial S}{\partial T}\bigg)_{\!p}=\frac{NT\Omega}{\Theta_DZ}.  %
}%
\end{array}
\end{equation}
In (\ref{58}), it is the notation $\displaystyle{Z\equiv 1-\frac{\Gamma\tau^2}{\big(1+\Gamma\big)}\frac{\Omega}{\varphi}}$. %

{\it Thermodynamic coefficients}. Similarly, using formulas
(\ref{54}),\,(\ref{55}), formulas can be obtained for the isobaric
volume expansion coefficient $\alpha_p$, isochoric pressure
coefficient $\beta_V$ and isothermal compressibility coefficient $\gamma_T$: %
\begin{equation} \label{59}
\begin{array}{ccc}
\displaystyle{%
  \beta_V=\frac{1}{p}\bigg(\frac{\partial p}{\partial T}\bigg)_{\!V}=\frac{T}{\Theta_D^2}\frac{\Omega}{\varphi},  %
}%
\end{array}
\end{equation}
\vspace{-4mm}
\begin{equation} \label{60}
\begin{array}{ccc}
\displaystyle{%
  \alpha_p=\frac{1}{V}\bigg(\frac{\partial V}{\partial T}\bigg)_{\!p}=\frac{T\Omega}{\Theta_D^2\big(1+\Gamma\big)\varphi Z},  %
}%
\end{array}
\end{equation}
\vspace{-4mm}
\begin{equation} \label{61}
\begin{array}{ccc}
\displaystyle{%
  \gamma_T=-\frac{1}{V}\bigg(\frac{\partial V}{\partial p}\bigg)_{\!T}=\frac{1}{\Gamma\big(1+\Gamma\big)n\Theta_D\varphi Z}.  %
}%
\end{array}
\end{equation}
The remaining thermodynamic coefficients, which can be constructed
from the quantities $T,\,S,\,p$ and $V$, are expressed in terms of
the coefficients and heat capacities given here \cite{R18}. The
difference in heat capacities, as required, is expressed through the
given above thermodynamic coefficients \cite{R17,R18}:
\begin{equation} \label{62}
\begin{array}{ccc}
\displaystyle{%
  C_p-C_V=\frac{VT}{\gamma_T}\,\alpha_p^2=N\frac{\Gamma}{\big(1+\Gamma\big)}\frac{T^3}{\Theta_D^3}\frac{\Omega^2}{\varphi Z}= %
  \frac{NT\Omega}{\Theta_D}\frac{\big(1-Z\big)}{Z}. %
}%
\end{array}
\end{equation}
Similar formulas for thermodynamic quantities of the normal state
according to the Debye theory are given for reference in Appendix C.

\vspace{0mm} %
\begin{figure}[b!]
\vspace{-0mm}  \hspace{0mm}
\includegraphics[width = 7.8cm]{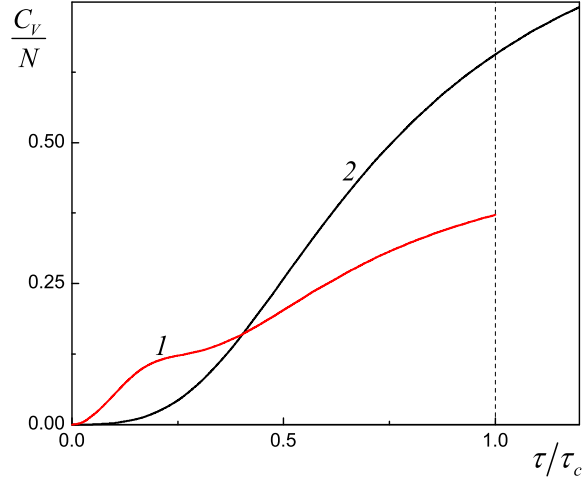} % 1.0\columnwidth
\vspace{-3mm} %
\caption{\label{fig05} %\hspace{10mm} %
The isochoric heat capacity in the asymmetric phase ({\it 1}) and
the Debye heat capacity ({\it 2}) at $|g|=0.7$.
}%
\end{figure}

\section{A jump of heat capacity}\vspace{-0mm} %
Above the phase transition temperature $\tau>\tau_c$, in the
self-consistent field model under consideration the phonon system is
described by the Debye model with the free energy $f_D(\tau)$
(\ref{40}). The transition to the phase-asymmetric state with pair
correlations is accompanied by an appearance of an order parameter
$\sigma$. This occurs at the temperature $\tau_c=T_c/\Theta_D$
determined by formula (\ref{42}). Let us consider the behavior of
the heat capacity near the transition temperature, assuming
$|\sigma|\ll 1$ and $\xi\approx 1-\frac{1}{2}|\sigma|^2-\frac{1}{8}|\sigma|^4$. %
Then the expansion of the free energy in the state with broken
symmetry at $\tau<\tau_c$ in powers of the order parameter takes the form %
\begin{equation} \label{63}
\begin{array}{ccc}
\displaystyle{%
  f(\tau)\approx f_D(\tau)+\alpha_0(\tau_c)\frac{(\tau-\tau_c)}{\tau_c}|\sigma|^2 + C(\tau_c)|\sigma|^4, %
}%
\end{array}
\end{equation}
where
\begin{equation} \label{64}
\begin{array}{ccc}
\displaystyle{%
  \alpha_0(\tau_c)=\frac{3}{8}\bigg[3g\Phi_2\big(\overline{\tau}_c\big){\rm cth}\bigg(\frac{1}{2\tau_c}\bigg)+ %
  \Phi_3\big(\overline{\tau}_c\big) + 2{\rm cth}\bigg(\frac{1}{2\tau_c}\bigg) \bigg], %
}%
\end{array}
\end{equation}
\vspace{-4mm}
\begin{equation} \label{65}
\begin{array}{ccc}
\displaystyle{%
  C(\tau_c)=\frac{9}{16}\Big[g\Phi_2\big(\overline{\tau}_c\big)+1 \Big]{\rm cth}\bigg(\frac{1}{2\tau_c}\bigg) + \frac{3}{64}\Phi_3\big(\overline{\tau}_c\big). %
}%
\end{array}
\end{equation}
Here, just like before, $\overline{\tau}_c=1/\tau_c$. From condition
$\partial f\big/\partial |\sigma|=0$ there follows the expression
for the equilibrium value of the order parameter
\begin{equation} \label{66}
\begin{array}{ccc}
\displaystyle{%
  |\sigma|^2=-\frac{\alpha_0(\tau_c)}{2C(\tau_c)}\frac{(\tau-\tau_c)}{\tau_c}. %
}%
\end{array}
\end{equation}
A substitution of (\ref{66}) into (\ref{63}) gives the expression
for the equilibrium free energy
\begin{equation} \label{67}
\begin{array}{ccc}
\displaystyle{%
  f(\tau)\approx f_D(\tau)-\frac{\alpha_0^2}{4C}\frac{(\tau-\tau_c)^2}{\tau_c^2}. %
}%
\end{array}
\end{equation}
It is usually assumed in the theory of phase transitions \cite{R17}
that the coefficients $\alpha_0(\tau_c)$ and $C(\tau_c)$ are
positive, so that the transition to the asymmetric phase is
accompanied by a decrease in the free energy. In the case under
consideration both coefficients $\alpha_0(\tau_c)$ and $C(\tau_c)$
are negative, so that, as noted above, the transition to the
asymmetric phase is accompanied by an increase in the free energy.
Formula (\ref{67}) makes it easy to obtain a formula for the jump in
the isochoric heat capacity during the transition from the symmetric
to the asymmetric phase:
\begin{equation} \label{68}
\begin{array}{ccc}
\displaystyle{%
  \Delta C_V=C_V(\tau_c+0)-C_V(\tau_c-0)=N\Theta_D\frac{\alpha_0^2}{2CT_c}. %
}%
\end{array}
\end{equation}
Jumps of other quantities can be found using well-known formulas
\cite{R17,R18}. The change in the heat capacity with temperature in
the asymmetric phase and the temperature dependence of the Debye
heat capacity are shown in Fig.\,5.

In the conventional theory of second-order phase transitions
\cite{R17}, the heat capacity increases by a jump upon transition
from the symmetric to the asymmetric phase. In this case, due to the
fact that $C(\tau_c)<0$, the heat capacity jump is negative. Thus,
upon transition from the normal Debye phase to the phase where there
exists a correlation of phonons with opposite momenta, the heat
capacity decreases by a jump.

\section{Region of low temperatures}\vspace{-0mm} %
Let us consider the thermodynamic quantities in the asymmetric phase
at low temperatures $T\ll\Theta_D$. It follows from formula
(\ref{41}) that at $T=0$ the parameter $\xi$ takes the finite value
$\xi_0=|g|$. Therefore, at $\tau\ll 1$ we have $\frac{\xi}{\tau}\gg
1$. The temperature dependence of $\xi$ at $\tau\ll 1$, up to
exponentially small terms, has the form
\begin{equation} \label{69}
\begin{array}{ccc}
\displaystyle{%
  \xi=\xi_0\big(1+a_3\tau^3-a_4\tau^4\big), %
}%
\end{array}
\end{equation}
where
\begin{equation} \label{70}
\begin{array}{ccc}
\displaystyle{%
  a_3=\frac{12\zeta(3)}{\xi_0}\bigg(\frac{5}{\xi_0^2}-3\bigg),  \qquad %
  a_4=\frac{8\pi^4}{15}\frac{1}{\xi_0^2}\bigg(\frac{5}{\xi_0^2}-4\bigg),  %
}%
\end{array}
\end{equation}
and $\zeta(3)=1.202$. The temperature dependence of the free energy
in the low-temperature limit is given by the formula
\begin{equation} \label{71}
\begin{array}{ccc}
\displaystyle{%
  f(\tau)=f_0-b_3\tau^3+b_4\tau^4, %
}%
\end{array}
\end{equation}
where
\begin{equation} \label{72}
\begin{array}{ccc}
\displaystyle{%
  f_0\equiv\frac{3}{16}\bigg(\xi_0+\frac{1}{\xi_0}\bigg),  \qquad %
  b_3\equiv\frac{9\zeta(3)}{2}\frac{1}{\xi_0^2}\bigg(\frac{1}{\xi_0^2}-1\bigg), \qquad   %
  b_4\equiv\frac{\pi^4}{5}\frac{1}{\xi_0^3}\bigg(\frac{1}{\xi_0^2}-\frac{4}{3}\bigg).   %
}%
\end{array}
\end{equation}
As we can see, in contrast to the corresponding dependence in the
Debye theory
\begin{equation} \label{73}
\begin{array}{ccc}
\displaystyle{%
  f_D(\tau)=\frac{3}{8}-\frac{\pi^4}{15}\,\tau^4,  %
}%
\end{array}
\end{equation}
in the case under consideration there is the main contribution to
the free energy of the cubic term with respect to temperature, and
the term of the fourth degree enters with the opposite sign. It
follows from (\ref{71}) that at $\tau\ll b_3\big/b_4$ both entropy
and heat capacity tend to zero at $T\rightarrow 0$ according to a
quadratic law with a cubic correction:
\begin{equation} \label{74}
\begin{array}{ccc}
\displaystyle{%
  \frac{S}{N}\approx 3b_3\tau^2-4b_4\tau^3, \qquad  %
  \frac{C_V}{N}\approx 6b_3\tau^2-12b_4\tau^3.  %
}%
\end{array}
\end{equation}
The decrease of these quantities in the low-temperature region is
slower than their cubic dependence according to the Debye theory.
The pressure and energy in this region are determined by the formula
\begin{equation} \label{75}
\begin{array}{ccc}
\displaystyle{%
  \frac{p}{\Gamma  n\Theta_D}=\frac{E}{N\Theta_D}=f_0+2b_3\tau^3-3b_4\tau^4.   %
}%
\end{array}
\end{equation}
The first term on the right side of (\ref{75}) describes the
contribution to the pressure and energy of zero-point lattice
vibrations.

\section{Region of existence of the asymmetric phase }\vspace{-0mm} %
The state of phonons with pair correlations considered in the
article can exist in a state of unstable thermodynamic equilibrium
only in those regions of temperatures and densities where the known
thermodynamic inequalities are satisfied \cite{R17}
\begin{equation} \label{76}
\begin{array}{ccc}
\displaystyle{%
  C_V>0, \qquad \big(\partial p\big/\partial V\big)_T<0.   %
}%
\end{array}
\end{equation}
The satisfiability of inequalities (\ref{76}) in the asymmetric
phase depends on the value of the interaction constant $|g|$, so
that they are satisfied under the condition
\begin{equation} \label{77}
\begin{array}{ccc}
\displaystyle{%
  0 < \Omega < \frac{\big(1+\Gamma\big)}{\Gamma}\frac{\varphi}{\tau^2}.   %
}%
\end{array}
\end{equation}
Condition (\ref{77}) also ensures the fulfillment of the inequality
$C_p-C_V>0$, since in this case $0<Z<1$. If inequalities (\ref{76})
are violated, then the system cannot exist even in a metastable
state.

Figure 6 shows the change in the isochoric heat capacity with
temperature for various values of the interaction constant. There
are two characteristic values $|g|_e=0.66$ and  $|g|_n=0.425$, at
which the behavior of the heat capacity changes qualitatively. In
the interval  $|g|_e<|g|<1$, the heat capacity increases
monotonically with increasing temperature (curve {\it 1} in
Fig.\,6). At $|g|=|g|_e$, an inflection point appears on the
temperature dependence of the heat capacity at $\tau_e/\tau_c=0.235$
(curve {\it 2}, Fig.\,6). In the range of values $|g|_n<|g|<|g|_e$,
the heat capacity remains positive, so that the stability conditions
(\ref{76}) are still satisfied, but it has a minimum (curve {\it 3},
Fig.\,6). At $|g|=|g|_n$ and temperature $\tau_n/\tau_c=0.151$, the
heat capacity turns to zero, and therefore the system loses
stability (curve {\it 4}, Fig.\,6). In the range of values
$0<|g|<|g|_n$, a temperature region arises where conditions
(\ref{76}) are not satisfied (curve {\it 5}, Fig.\,6). As the
interaction constant decreases, this region of instability expands.
\vspace{0mm} %
\begin{figure}[h!]
\vspace{-0mm}  \hspace{0mm}
\includegraphics[width = 7.8cm]{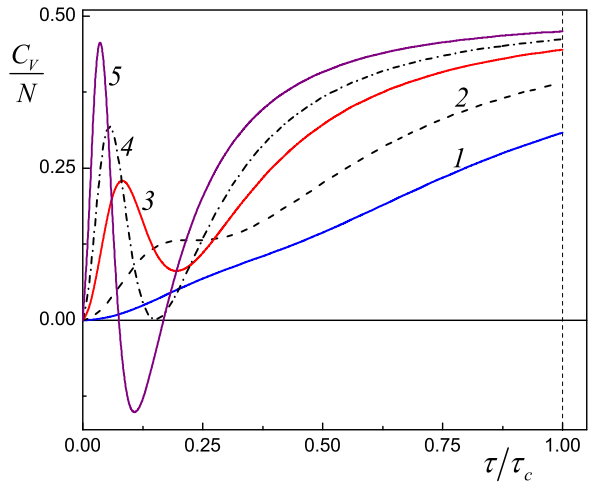} % 1.0\columnwidth
\vspace{-3mm} %
\caption{\label{fig06} %\hspace{10mm} %
The isochoric heat capacity in the asymmetric phase for different
values of the interaction constant: %
{\it 1}) $|g|=0.8$; %
{\it 2}) $|g|=0.66$; %
{\it 3}) $|g|=0.5$; %
{\it 4}) $|g|=0.425$; %
{\it 5}) $|g|=0.35$. %
}%
\end{figure}

As noted above, the critical temperature increases as the magnitude
of the interaction decreases (Fig.\,1), but, as we see, there
narrows the temperature range where the metastable existence of a
new phase is possible. Thus, the transition temperature is not
likely to be high in reality.

\section{Discussion. Conclusions}\vspace{-0mm} %
It is shown in the paper that a gas of phonons, between which there
is an attraction, can be in the unstable equilibrium state with
broken phase symmetry in which there exist pair correlations between
phonons with opposite momenta. The transition to such a state from
the normal state, which is described by the Debye theory, can occur
at a certain critical temperature as a result of the second-order
phase transition. The critical temperature is determined by the
value of the interaction constant, and it turns out that the
critical temperature increases with the weakening of the interaction
(Fig.\,1). However, with a decrease in the modulus of the
interaction constant and an increase in the critical temperature,
there expands the temperature range in which the thermodynamic
stability of the asymmetric phase is completely lost. In contrast to
conventional equilibrium phase transitions, in this case it turns
out that the free energy of the asymmetric state is higher than the
energy of the normal state. This leads to the fact that such a state
turns out to be metastable, and upon transition from the normal
state to the asymmetric phase the heat capacity decreases by a jump.

Qualitatively, the situation considered in the paper can be
explained by a simple example of a physical pendulum, which can be
in two extreme states. In the lower position its energy is minimal
and this state is stable. In the upper position the energy of the
pendulum is maximum, its position is unstable and it can be
withdrawn from it at the slightest fluctuation. If, however, the
friction in the suspension of the pendulum is large, then it can
remain in the upper metastable state for quite a long time.

Similarly, if the relaxation time of a state of a solid with
pairwise phonon correlations turns out to be long, then such a state
may be accessible for observation. The question of how the system
can be transferred to the considered unstable state requires a
separate study. There are considerations that this can be done with
the help of external high-frequency forces. It is also possible that
the new phase can be identified by its influence on the propagation
and absorption of sound \cite{R19}. The considered phase may turn
out to be more stable under inhomogeneous conditions, for example,
near boundaries. Theoretically, such states can also be studied
within the framework of the generalized Debye model \cite{R20}.

In this paper, the problem was considered in the scalar
approximation without taking into account phonon polarizations,
under the assumption that the speed of all phonons is the same.
However, already in an isotropic medium there are two elastic moduli
and, consequently, two types of phonons -- longitudinal and
transverse. This circumstance must be taken into account in a more
realistic model \cite{R21}. In addition, nonlinear effects in the
considered model were taken into account in a minimal way by
introducing the interaction of phonons with opposite momenta into
the Hamiltonian (\ref{01}). In real crystals, a significant role can
be played by nonlinear effects described by elastic moduli of the
third and fourth order \cite{R12,R22}. Nonlinear effects play a
particularly important role in crystals in which quantum
regularities are manifested \cite{R23}. In our calculations we
neglected the contribution of the correlation Hamiltonian (\ref{04})
to the total energy. Accounting for the correlation energy can also
lead to the stabilization of the considered state. Thus, the model
of a crystal with correlations of phonons with opposite momenta
needs further development and generalization. Phase symmetry
breaking in multiparticle systems is accompanied by the
manifestation of superfluid properties \cite{R24,R25,R26,R27}. In
the considered model, the phase transition is also accompanied by
the breaking of the phase symmetry. In this connection, as well as
in connection with experimental studies of quantum crystals
\cite{R10}, the question of the manifestation of superfluidity
effects in phonon systems is of considerable interest.

The author is grateful to A.A.\,Soroka for help with numerical
calculations.

\appendix

\section{Debye functions with integer index $n\ge 1$ and related functions} %
{\it Definitions}:
\begin{equation} \label{A1}
\begin{array}{ccc}
\displaystyle{%
  D_n(x)=\frac{n}{x^n}\int_0^x\frac{z^ndz}{e^z-1},  \qquad (n\ge 1),   %
}%
\end{array}
\end{equation}
\vspace{-4mm} %
\begin{equation} \label{A2}
\begin{array}{ccc}
\displaystyle{%
  \Phi_n(x)=1+\frac{2(1+n)}{n}\frac{D_n(x)}{x}.   %
}%
\end{array}
\end{equation}
Functions (\ref{A1}) can be represented as
\begin{equation} \label{A3}
\begin{array}{ccc}
\displaystyle{%
  D_n(x)=\frac{n}{x^n}\bigg[n!\zeta(n+1)-\sum_{m=0}^\infty\int_x^\infty e^{-(m+1)z}z^ndz\bigg],  %
}%
\end{array}
\end{equation}
$\zeta(n+1)$ is the Riemann zeta function. The functions with
$n=2,3$ are used in this paper, so that
\begin{equation} \label{A4}
\begin{array}{ccc}
\displaystyle{%
  \Phi_2(x)=1+3\frac{D_2(x)}{x}, \qquad  \Phi_3(x)=1+\frac{8}{3}\frac{D_3(x)}{x}.   %
}%
\end{array}
\end{equation}

{\it Large arguments} $x\gg 1$: %
Up to exponentially small terms:
\begin{equation} \label{A5}
\begin{array}{ccc}
\displaystyle{%
  D_n(x)\approx\frac{n\!\cdot\!n!}{x^n}\zeta(n+1), \qquad %
  D_2(x)\approx\frac{4}{x^2}\zeta(3), \qquad D_3(x)\approx\frac{\pi^4}{5x^3},   %
}\vspace{3mm}\\ %
\displaystyle{%
  \Phi_n(x)\approx 1+\frac{2(1+n)\!\cdot\!n!}{x^{n+1}}\zeta(n+1), \qquad %
  \Phi_2(x)\approx 1+\frac{12}{x^3}\zeta(3), \qquad %
  \Phi_3(x)\approx 1+\frac{8\pi^4}{15x^4}. %
}%
\end{array}
\end{equation}

{\it Small arguments} $x\ll 1$: %
\begin{equation} \label{A6}
\begin{array}{ccc}
\displaystyle{%
  D_n(x)\approx 1-\frac{n}{2(n+1)}\,x+\frac{n}{12(n+2)}\,x^2, \quad
  D_2(x)\approx 1-\frac{x}{3}+\frac{x^2}{24}, \quad    %
  D_3(x)\approx 1-\frac{3}{8}\,x+\frac{x^2}{20}, \   %
}\vspace{3mm}\\ %
\displaystyle{%
  \Phi_n(x)\approx\frac{2(n+1)}{nx}+\frac{(n+1)}{6(n+2)}\,x, \quad %
  \Phi_2(x)\approx\frac{3}{x}+\frac{x}{8}, \quad %
  \Phi_3(x)\approx\frac{8}{3x}+\frac{2}{15}\,x.
}%
\end{array}
\end{equation}

{\it Derivatives} (used designations are $dD_n(x)\big/dx\equiv
\dot{D}_n(x)$, $d^2D_n(x)\big/dx^2\equiv \ddot{D}_n(x)$ and so on): %
\begin{equation} \label{A7}
\begin{array}{ccc}
\displaystyle{%
  \dot{D}_n(x)=\frac{n}{e^x-1}-\frac{n}{x}D_n(x),  %
}\vspace{3mm}\\ %
\displaystyle{%
  \dot{D}_2(x)=\frac{2}{e^x-1}-\frac{2}{x}D_2(x),  \qquad %
  \dot{D}_3(x)=\frac{3}{e^x-1}-\frac{3}{x}D_3(x);  %
}%
\end{array}
\end{equation}
%%%
\begin{equation} \label{A8}
\begin{array}{ccc}
\displaystyle{%
  \dot{\Phi}_n(x)=\frac{2(n+1)}{x(e^x-1)}+\frac{(n+1)}{x}-\frac{(n+1)}{x}\,\Phi_n(x)=\frac{(n+1)}{x}\bigg[{\rm cth}\frac{x}{2}-\Phi_n(x)\bigg],  %
}\vspace{3mm}\\ %
\displaystyle{%
  \dot{\Phi}_2(x)=\frac{6}{x(e^x-1)}+\frac{3}{x}-\frac{3}{x}\,\Phi_2(x)=\frac{3}{x}\bigg[{\rm cth}\frac{x}{2}-\Phi_2(x)\bigg],  %
}\vspace{3mm}\\ %
\displaystyle{%
  \dot{\Phi}_3(x)=\frac{8}{x(e^x-1)}+\frac{4}{x}-\frac{4}{x}\,\Phi_3(x)=\frac{4}{x}\bigg[{\rm cth}\frac{x}{2}-\Phi_3(x)\bigg].  %
}%
\end{array}
\end{equation}
%%%
\begin{equation} \label{A9}
\begin{array}{ccc}
\displaystyle{%
  \ddot{\Phi}_n(x)=-\frac{2(n+1)}{x(e^x-1)}\bigg(1+\frac{(n+2)}{x}\bigg)-\frac{2(n+1)}{x(e^x-1)^2}  %
  -\frac{(n+1)(n+2)}{x^2}+\frac{(n+1)(n+2)}{x^2}\,\Phi_n(x), %
}\vspace{3mm}\\ %
\displaystyle{%
  \ddot{\Phi}_2(x)=-\frac{6}{x(e^x-1)}\bigg(1+\frac{4}{x}\bigg)-\frac{6}{x(e^x-1)^2}  %
  -\frac{12}{x^2}+\frac{12}{x^2}\,\Phi_2(x), %
}\vspace{3mm}\\ %
\displaystyle{%
  \ddot{\Phi}_3(x)=-\frac{8}{x(e^x-1)}\bigg(1+\frac{5}{x}\bigg)-\frac{8}{x(e^x-1)^2}  %
  -\frac{20}{x^2}+\frac{20}{x^2}\,\Phi_3(x). %
}%
\end{array}
\end{equation}

\section{The function $f(\xi,\tau)$ (\ref{37}) and its derivatives:} \vspace{-6.7mm} %
\begin{equation} \label{B1}
\begin{array}{ccc}
\displaystyle{%
  f(\xi,\tau)=\frac{\xi}{2}+\frac{3}{16}\,g\big(\xi^{-2}-1\big)\Phi_2^2(\xi\overline{\tau}) %
  +\frac{3}{8}\Big(\xi^{-1}-\frac{4}{3}\xi\Big)\Phi_3(\xi\overline{\tau}) +\tau\ln\!\big(1-e^{-\xi\overline{\tau}}\big),   %
}%
\end{array}
\end{equation}
%%%
\begin{equation} \label{B2}
\begin{array}{ccc}
\displaystyle{%
 \frac{\partial f}{\partial\xi}\equiv\psi(\xi,\tau)=\frac{1}{2}\,{\rm cth}\frac{\xi}{2\tau} %
 -\frac{3g}{8}\xi^{-3}\Phi_2^2(\xi\overline{\tau}) %
 +\frac{3g}{8\tau}\big(\xi^{-2}-1\big)\Phi_2(\xi\overline{\tau})\dot{\Phi}_2(\xi\overline{\tau})\,- %
}\vspace{3mm}\\ %
\displaystyle{%
  -\frac{3}{8}\Big(\xi^{-2}+\frac{4}{3}\Big)\Phi_3(\xi\overline{\tau}) %
  +\frac{3}{8\tau}\Big(\xi^{-1}-\frac{4}{3}\xi\Big)\dot{\Phi}_3(\xi\overline{\tau}), %
}%
\end{array}
\end{equation}
%%%
\begin{equation} \label{B3}
\begin{array}{ccc}
\displaystyle{%
 \frac{\partial f}{\partial\tau}\equiv -s(\xi,\tau)=-\frac{3g}{8}\frac{\xi}{\tau^2}\big(\xi^{-2}-1\big)\Phi_2(\xi\overline{\tau})\dot{\Phi}_2(\xi\overline{\tau}) %
 -\frac{3}{8}\frac{\xi}{\tau^2}\Big(\xi^{-1}-\frac{4}{3}\xi\Big)\dot{\Phi}_3(\xi\overline{\tau})\,+ %
}\vspace{3mm}\\ %
\displaystyle{%
  +\ln\!\big(1-e^{-\xi\overline{\tau}}\big)-\frac{\xi}{\tau}\frac{1}{e^{\xi\overline{\tau}}-1}, %
}%
\end{array}
\end{equation}
%%%
\begin{equation} \label{B4}
\begin{array}{ccc}
\displaystyle{%
 \varphi(\xi,\tau)\equiv f-\tau\frac{\partial f}{\partial\tau}=\frac{\xi}{2}\,{\rm cth}\frac{\xi}{2\tau}  %
 +\frac{3g}{16}\big(\xi^{-2}-1\big)\Phi_2(\xi\overline{\tau})\Big[\Phi_2(\xi\overline{\tau})+2\frac{\xi}{\tau}\dot{\Phi}_2(\xi\overline{\tau})\Big]+ %
}\vspace{3mm}\\ %
\displaystyle{%
  +\frac{3}{8}\Big(\xi^{-1}-\frac{4}{3}\xi\Big)\Big[\Phi_3(\xi\overline{\tau})+\frac{\xi}{\tau}\dot{\Phi}_3(\xi\overline{\tau})\Big], %
}%
\end{array}
\end{equation}
%%%
\begin{equation} \label{B5}
\begin{array}{ccc}
\displaystyle{%
 \frac{\partial^2f}{\partial \xi^2}=\frac{1}{4\tau}\bigg[1-{\rm cth}^2\frac{\xi}{2\tau}\bigg] +  %
}\vspace{3mm}\\ %
\displaystyle{%
  +\frac{3g}{8}\bigg[\frac{3}{\xi^4}\Phi_2^2(\xi\overline{\tau})-\frac{4}{\xi^3\tau}\Phi_2(\xi\overline{\tau})\dot{\Phi}_2(\xi\overline{\tau}) %
  +\frac{1}{\tau^2}\big(\xi^{-2}-1\big)\dot{\Phi}_2^2(\xi\overline{\tau}) %
  +\frac{1}{\tau^2}\big(\xi^{-2}-1\big)\Phi_2(\xi\overline{\tau})\ddot{\Phi}_2(\xi\overline{\tau})\bigg] + %
}\vspace{3mm}\\ %
\displaystyle{%
  +\frac{3}{4\xi^3}\Phi_3(\xi\overline{\tau})-\frac{3}{4\tau}\Big(\xi^{-2}+\frac{4}{3}\Big)\dot{\Phi}_3(\xi\overline{\tau}) %
  +\frac{3}{8\tau^2}\Big(\xi^{-1}-\frac{4}{3}\xi\Big)\ddot{\Phi}_3(\xi\overline{\tau}), %
}%
\end{array}
\end{equation}
%%%
\begin{equation} \label{B6}
\begin{array}{ccc}
\displaystyle{%
 \frac{\partial^2f}{\partial \tau^2}=\frac{3g}{4}\frac{\xi}{\tau^3}\big(\xi^{-2}-1\big)\Phi_2(\xi\overline{\tau})\dot{\Phi}_2(\xi\overline{\tau})  %
 +\frac{3g}{8}\frac{\xi^2}{\tau^4}\big(\xi^{-2}-1\big)\Big[\dot{\Phi}_2^2(\xi\overline{\tau})+\Phi_2(\xi\overline{\tau})\ddot{\Phi}_2(\xi\overline{\tau})\Big]+  %
}\vspace{3mm}\\ %
\displaystyle{%
  +\frac{3}{4}\frac{\xi}{\tau^3}\Big(\xi^{-1}-\frac{4}{3}\xi\Big)\dot{\Phi}_3(\xi\overline{\tau}) %
  +\frac{3}{8}\frac{\xi^2}{\tau^4}\Big(\xi^{-1}-\frac{4}{3}\xi\Big)\ddot{\Phi}_3(\xi\overline{\tau}) %
  -\frac{\xi^2}{\tau^3}\frac{e^{\xi\overline{\tau}}}{\big(e^{\xi\overline{\tau}}-1\big)^2}, %
}%
\end{array}
\end{equation}
%%%
\begin{equation} \label{B7}
\begin{array}{ccc}
\displaystyle{%
 \frac{\partial^2f}{\partial\xi\partial\tau}= \frac{3g}{8}\frac{1}{\tau^2}\big(\xi^{-2}+1\big)\Phi_2(\xi\overline{\tau})\dot{\Phi}_2(\xi\overline{\tau}) %
-\frac{3g}{8}\frac{\xi}{\tau^3}\big(\xi^{-2}-1\big)\Big[\dot{\Phi}_2^2(\xi\overline{\tau})+\Phi_2(\xi\overline{\tau})\ddot{\Phi}_2(\xi\overline{\tau})\Big]+  %
}\vspace{3mm}\\ %
\displaystyle{%
  +\frac{\xi}{\tau^2}\dot{\Phi}_3(\xi\overline{\tau})-\frac{3}{8}\frac{\xi}{\tau^3}\Big(\xi^{-1}-\frac{4}{3}\xi\Big)\ddot{\Phi}_3(\xi\overline{\tau})  %
  +\frac{\xi}{\tau^2}\frac{e^{\xi\overline{\tau}}}{\big(e^{\xi\overline{\tau}}-1\big)^2}. %
}%
\end{array}
\end{equation}

\section{Thermodynamic functions in the Debye free phonon model} %
{\it The free energy}:
\begin{equation} \label{C1}
\begin{array}{ccc}
\displaystyle{%
  \frac{F_D}{\Theta_DN}\equiv f_D(\tau)=\frac{1}{2}-\frac{1}{8}\Phi_3(\overline{\tau})+\tau\ln\!\big(1-e^{-\overline{\tau}}\big), %
}%
\end{array}
\end{equation}
the entropy
\begin{equation} \label{C2}
\begin{array}{ccc}
\displaystyle{%
  \frac{S}{N}=-\frac{\partial f_D}{\partial\tau}=\frac{4}{3}D_3(\overline{\tau})-\ln\!\big(1-e^{-\overline{\tau}}\big), %
}%
\end{array}
\end{equation}
the pressure
\begin{equation} \label{C3}
\begin{array}{ccc}
\displaystyle{%
  p=\Gamma n\Theta_D\bigg(f_D-\tau\frac{\partial f_D}{\partial\tau}\bigg)=\frac{3}{8}\Gamma n\Theta_D\Phi_3(\overline{\tau}), %
}%
\end{array}
\end{equation}
the energy
\begin{equation} \label{C4}
\begin{array}{ccc}
\displaystyle{%
  E_D=N\Theta_D\bigg(f_D-\tau\frac{\partial f_D}{\partial\tau}\bigg)=\frac{3}{8}N\Theta_D\Phi_3(\overline{\tau})=\frac{pV}{\Gamma}, %
}%
\end{array}
\end{equation}
the phonon number
\begin{equation} \label{C5}
\begin{array}{ccc}
\displaystyle{%
  N_{ph}^{(D)}=\frac{3}{2}N\tau D_2(\overline{\tau}). %
}%
\end{array}
\end{equation}
The entropy and pressure differentials:
\begin{equation} \label{C6}
\begin{array}{ccc}
\displaystyle{%
  \frac{dS}{N}=\tau\Omega_0\bigg(\frac{dT}{T}+\Gamma\frac{dV}{V}\bigg),  %
}%
\end{array}
\end{equation}
\vspace{-4mm} %
\begin{equation} \label{C7}
\begin{array}{ccc}
\displaystyle{%
  dp=\Gamma n\Omega_0T\frac{dT}{\Theta_D}+\Gamma\big(1+\Gamma\big)n\Theta_D\varphi_0 %
  \bigg[\frac{\Gamma}{\big(1+\Gamma\big)}\frac{\Omega_0\tau^2}{\varphi_0}-1\bigg]\frac{dV}{V},  %
}%
\end{array}
\end{equation}
where
\begin{equation} \label{C8}
\begin{array}{ccc}
\displaystyle{%
  \Omega_0\equiv\frac{1}{\tau^2}\bigg[4\tau D_3(\overline{\tau})-\frac{3}{e^{\overline{\tau}}-1}\bigg]  %
  =-\frac{3}{8}\frac{\dot{\Phi}_3(\overline{\tau})}{\tau^3}, \qquad %
  \varphi_0\equiv\frac{3}{8}\Phi_3(\overline{\tau}). %
}%
\end{array}
\end{equation}
The formulas for the heat capacities and thermodynamic coefficients
coincide with formulas (\ref{56})\,--\,(\ref{61}) for the asymmetric
phase, if we make the following substitutions in them
\begin{equation} \nonumber
\begin{array}{ccc}
\displaystyle{%
  \Omega\rightarrow\Omega_0, \quad \varphi\rightarrow\varphi_0, \quad %
  Z\rightarrow Z_0\equiv 1-\frac{\Gamma}{\big(1+\Gamma\big)}\frac{\Omega_0\tau^2}{\varphi_0},  %
}%
\end{array}
\end{equation}
so that we obtain
\begin{equation} \label{C9}
\begin{array}{ccc}
\displaystyle{%
  C_V=\frac{NT\Omega_0}{\Theta_D}, \qquad %
  C_p=\frac{NT\Omega_0}{\Theta_D}\bigg(1+\frac{\Gamma}{1+\Gamma}\frac{\Omega_0\tau^2}{\varphi_0Z_0}\bigg), %
}\vspace{3mm}\\ %
\displaystyle{%
   C_p-C_V=\frac{VT}{\gamma_T}\alpha_p^2=N\frac{\Gamma}{\big(1+\Gamma\big)}\frac{T^3}{\Theta_D^3}\frac{\Omega_0^2}{\varphi_0Z_0},  %
}%
\end{array}
\end{equation}
and also
\begin{equation} \label{10}
\begin{array}{ccc}
\displaystyle{%
  \beta_V=\frac{T}{\Theta_D^2}\frac{\Omega_0}{\varphi_0}, \qquad %
  \alpha_p=\frac{T\Omega_0}{\Theta_D^2\big(1+\Gamma\big)\varphi_0Z_0}, \qquad %
  \gamma_T=\frac{1}{\Gamma\big(1+\Gamma\big)n\Theta_D\varphi_0Z_0}. %
}%
\end{array}
\end{equation}
The stability condition (\ref{77}) for the Debye state is always
satisfied.

%\vspace{20mm} %

\end{document}